\documentclass[12pt]{iopart}
\usepackage{graphicx,upgreek,nowidow,times,epstopdf,color}

\usepackage{iopams}  

\expandafter\let\csname equation*\endcsname\relax

\expandafter\let\csname endequation*\endcsname\relax

\usepackage{amsmath,amsthm,bm}

\newcommand{\ket}[1]{\left|{#1}\right>}
\newcommand{\bra}[1]{\left<{#1}\right|}

\newcommand{\opinner}[3]{\left<{#1}\vphantom{#1}\vphantom{#3}\right|{#2}\left|{#3}\vphantom{#1}\vphantom{#3}\right>}
\newcommand{\rvec}[1]{\pmb{#1}}
\newcommand{\dyadic}[1]{\pmb{#1}}

\newcommand{\D}{\mathrm{d}}
\newcommand{\I}{\mathrm{i}}
\newcommand{\TP}[1]{{#1}^\mathrm{\,\textsc{t}}}
\newcommand{\E}[1]{\mathrm{e}^{\mbox{\footnotesize$#1$}}}

\newcommand{\DET}[1]{\mathrm{Det}\!\left\{#1\right\}}

\newcommand{\ML}[1]{\widehat{#1}_\textsc{ml}}

\newcommand{\norm}[1]{\left|\!\left|#1\right|\!\right|}
\newcommand{\FML}{\dyadic{F}_\textsc{ml}}

\newcommand{\MEAN}[2]{\mathrm{E}_{#1}\!\left[#2\right]}
\renewcommand{\MSE}[1]{\mathrm{MSE}\!\left(#1\right)}
\newcommand{\RSE}[2]{\mathrm{RSE}_{#1}\!\left(#2\right)}
\newcommand{\MRSEc}[1]{\mathrm{MRSE}^{\text{(cred)}}\!\left(#1\right)}
\newcommand{\MRSEp}[1]{\mathrm{MRSE}^{\text{(plaus)}}\!\left(#1\right)}

\newcommand{\goodline}{
\begin{center}
	\rule{0.1\textwidth}{0.5pt}\\[-2.9ex]
	\rule{0.2\textwidth}{0.5pt}\\[-2.9ex]
	\rule{0.3\textwidth}{0.5pt}\\[-2.9ex]
	\rule{0.35\textwidth}{0.5pt}\\[-2.9ex]	
	\rule{0.3\textwidth}{0.5pt}\\[-2.9ex]
	\rule{0.2\textwidth}{0.5pt}\\[-2.9ex]
	\rule{0.1\textwidth}{0.5pt}
\end{center}}

\begin{document}

\title{Bayesian error regions in quantum estimation II: region accuracy and adaptive methods}

\author{Changhun Oh$^1$, Yong Siah Teo$^{1,2}$, and Hyunseok Jeong$^1$}

\address{$^1$ Center for Macroscopic Quantum Control, Seoul National University, 08826 Seoul, South Korea}
\address{$^2$ BK21 Frontier Physics Research Division, Seoul National University, 08826 Seoul, South Korea}
\ead{v55ohv@snu.ac.kr and ys\_teo@snu.ac.kr}
\vspace{10pt}

\begin{abstract}
Bayesian error analysis paves the way to the construction of credible and plausible error regions for a point estimator obtained from a given dataset. We introduce the concept of region accuracy for error regions (a generalization of the point-estimator mean squared-error) to quantify the average statistical accuracy of all region points with respect to the unknown true parameter. We show that the increase in region accuracy is closely related to the Bayesian-region dual operations in \cite{Shang:2013cc}. Next with only the given dataset as viable evidence, we establish various adaptive methods to maximize the region accuracy relative to the true parameter subject to the type of reported Bayesian region for a given point estimator. We highlight the performance of these adaptive methods by comparing them with nonadaptive procedures in three quantum-parameter estimation examples. The results of and mechanisms behind the adaptive schemes can be understood as the region analog of adaptive approaches to achieving the quantum Cram{\'e}r--Rao bound for point estimators.
\end{abstract}

%
\vspace{2pc}

\section{Introduction}

Error estimation for a given particular point estimator of an unknown parameter constitutes an important component in quantum estimation. The assigned error interval (or region) for the estimator conveys error information about the measured data that propagates to other physical quantities predicted with this estimator. 

The preceding companion article~\cite{BR:part1} discussed asymptotic techniques for constructing Bayesian regions for the maximum-likelihood (ML) estimator. Such a Bayesian region annotates the estimator with credibility that it lies in this region of a given size. As a result, we see that this construction contains statistical elements from two principal schools of thought. If one is pedantic about labeling these elements, one might say that the concept of an unknown, but fixed, parameter is that of a frequentist, whereas terminologies like size and credibility for a distribution of parameters belong to viewpoints of a Bayesian~\cite{Efron:1986aa,Cousins:1995aa,Suess:2017np}.

In this article, rather than distinguishing between these two schools, we shall understand the underlying meaning of statistical accuracy in the construction of Bayesian regions that is based on elements from these two camps. As a means to eradicate unnecessary confusion, we note here that in relation to Bayesian statistics, an important point estimator of interest is the average of the product of both the parameter and its posterior distribution over the entire parameter space---the Bayesian mean estimator. For this point estimator, concepts of statistical accuracy exist~\cite{Blume-Kohout:2010aa,Efron:2015aa}. 

In our context however, the relation between statistical accuracy for the unknown true parameter and aspects of Bayesian regions comes in a different flavor. We are still interested in a frequentist accuracy for the true parameter of interest, just as much as an observer is interested in preparing a quantum source in a particular state, for instance. On the other hand, since we are dealing with error regions, which are sets of points, we now speak of the \emph{region accuracy}, that is the average accuracy of \emph{all} points in the region relative to the true parameter. In the limit of zero region size, the region accuracy becomes the usual point-estimator accuracy. After a review in Sec.~\ref{sec:basics} on Bayesian regions, we shall see in Sec.~\ref{sec:statacc} that this notion of region accuracy is intimately connected to the dual operations~\cite{BR:part1,Shang:2013cc} of fixing the region size while increasing credibility, or fixing the credibility---both actions \emph{tend to} increase the region accuracy, and this tendency becomes \emph{exact} in single-parameter quantum estimation.

In Sec.~\ref{sec:adap} we will proceed to develop operational schemes to maximize region accuracy by either adaptively optimizing over all credible regions of fixed size/credibility, or over all plausible regions. The adaptive schemes require solely the collected data and parameter dimension, and are in fact region analogs of adaptively attaining the quantum Cram{\'e}r--Rao bound for point estimators~\cite{Yuen:1973aa,Helstrom:1974aa,Braunstein:1994sd}. These schemes will be applied to three examples in quantum estimation that can be categorized under quantum metrology and Gaussian-state characterization. All symbols and notations from \cite{BR:part1} are carried over to this article. The prior distribution for the true parameter shall again be taken to be the uniform primitive distribution in the parameter space.

\section{Brief review on Bayesian regions}
\label{sec:basics}

For the purpose of laying out the foundations for subsequent discussion on region accuracy and adaptive quantum estimation, we state the key properties of a Bayesian credible region $\mathcal{R}=\mathcal{R}_\lambda$ that is characterized by $0\leq\lambda\leq1$ with an isolikelihood boundary. The size and credibility of $\mathcal{R}_\lambda$ are defined in (1) of \cite{BR:part1}. 

From hereon, we shall focus~(see later Sec.~\ref{subsec:cases23}) on the situation where the true parameter $\rvec{r}\notin\partial\mathcal{R}$, so that for a sufficiently large data sample size $N$, the error region $\mathcal{R}$ for all interesting values of $\lambda$ has boundary $\partial\mathcal{R}\cap\partial\mathcal{R}_0=\emptyset$ (Case~1 in \cite{BR:part1}). For the case in point, we reiterate the relevant expressions
\begin{eqnarray}
s_\lambda&=&\,\frac{V_d}{V_{\mathcal{R}_0}}(-2\log\lambda)^{d/2}\,\DET{\FML}^{-1/2}\,,\nonumber\\
c_\lambda&=&\,1-\frac{\Gamma(d/2, -\log\lambda)}{(d/2-1)!}\,,\nonumber\\
\lambda_{\rm{crit}}&=&\,\sqrt{\DET{2\pi\,\FML^{-1}}}/V_{\mathcal{R}_0}\,,
\label{eq:rem_case1}
\end{eqnarray}
for the size, credibility and the critical $\lambda=\lambda_{\rm{crit}}$ that defines the plausible region---the credible region that contains \emph{all plausible parameters and nothing else}. Based on these expressions, we can obtain the simple relation
\begin{eqnarray}
s_\lambda=\frac{V_d}{V_{\mathcal{R}_0}}\left[2\,\Gamma^{-1}_{d/2}\left(1-c_\lambda\right)\right]^{d/2}\,\DET{\FML}^{-1/2}
\label{eq:sc}
\end{eqnarray}
between $s_\lambda$ and $c_\lambda$. These analytical results apply to the uniform primitive prior with respect to the parameter $\rvec{r}$. More explicitly, for $\rvec{r}\,\widehat{=}\,\TP{(r_1\,\,r_2\,\,\ldots\,\,r_d)}$, the integral measure $(\D\,\rvec{r})=(\prod_j \D\,r_j)/V_{\mathcal{R}_0}$.

\section{Region accuracy and its connections with the dual region operations}
\label{sec:statacc}

\subsection{General formalism of the region accuracy}

Suppose that after collecting the experimental data $\mathbb{D}$, the ML estimator $\ML{\rvec{r}}$ is computed over the parameter space $\mathcal{R}_0$. Then the usual mean squared error~(MSE) for this point estimator relative to the true parameter $\rvec{r}$,
\begin{eqnarray}
\mathrm{MSE}(\rvec{r})\equiv\MEAN{}{\norm{\ML{\rvec{r}}-\rvec{r}}^2}\,,
\end{eqnarray}
measures the average statistical accuracy of $\ML{\rvec{r}}$ over all possible data $\mathbb{D}$. It is known that if Case~1 applies, then for sufficiently large $N$ the ML estimator will ultimately be unbiased ($\MEAN{}{\ML{\rvec{r}}}=\rvec{r}$) and so $\MSE{\rvec{r}}\rightarrow\Tr\{\dyadic{F}^{-1}\}$ approaches the Cram{\'e}r--Rao bound that is defined by the Fisher information $\dyadic{F}$ for $\rvec{r}$.

We may generalize this description of accuracy using the language of Bayesian analysis on the ML estimator $\ML{\rvec{r}}$. Since the object in this analysis is the Bayesian region $\mathcal{R}$, it is natural to introduce the \emph{region squared error} (RSE)
\begin{eqnarray}
\RSE{}{\rvec{r}}=\dfrac{\int_{\mathcal{R}}(\D\,\rvec{r}')\norm{\rvec{r}'-\rvec{r}}^2}{\int_{\mathcal{R}}(\D\,\rvec{r}')}
\end{eqnarray}
that measures the region accuracy relative to $\rvec{r}$, or the average accuracy of all the points in $\mathcal{R}$, where $(\D\,\rvec{r}')$ is the normalized integral measure as defined in \cite{BR:part1}. It is easy to see that when $\mathcal{R}=\mathcal{R}_{\lambda=1}=\{\ML{\rvec{r}}\}$, we return to $\RSE{}{\rvec{r}}=\MSE{\rvec{r}}$ since for any function $f(\rvec{r})$,
\begin{eqnarray}
\qquad\,\,\,\dfrac{\int_{\mathcal{R}_{\lambda\rightarrow1}}(\D\,\rvec{r}')\,f(\rvec{r}')}{\int_{\mathcal{R}_{\lambda\rightarrow1}}(\D\,\rvec{r}')}&=&\,f(\ML{\rvec{r}})+\dfrac{\int_{\mathcal{R}_{\lambda\rightarrow1}}(\D\,\rvec{r}')\,(\rvec{r}'-\ML{\rvec{r}})}{\int_{\mathcal{R}_{\lambda\rightarrow1}}(\D\,\rvec{r}')}\bm{\cdot}\bm{\partial}_\textsc{ml}f(\ML{\rvec{r}})\,,\nonumber\\ \norm{\dfrac{\int_{\mathcal{R}_{\lambda\rightarrow1}}(\D\,\rvec{r}')\,(\rvec{r}'-\ML{\rvec{r}})}{\int_{\mathcal{R}_{\lambda\rightarrow1}}(\D\,\rvec{r}')}}&\leq&\,\max_{\rvec{r}'\in\mathcal{R}_{\lambda\rightarrow1}}\norm{\rvec{r}'-\ML{\rvec{r}}}=0\,.
\end{eqnarray}
To analyze the average region accuracy over all possible data for the error regions, we may adopt the \emph{mean region squared error} (MRSE)
\begin{eqnarray}
\mathrm{MRSE}(\rvec{r})\equiv\MEAN{}{\RSE{}{\rvec{r}}}=\MEAN{}{\dfrac{\int_{\mathcal{R}}(\D\,\rvec{r}')\norm{\rvec{r}'-\rvec{r}}^2}{\int_{\mathcal{R}}(\D\,\rvec{r}')}}\,.
\end{eqnarray}
Statistically, the MRSE is a collective error feature of the ML point estimator $\ML{\rvec{r}}$ \emph{and} its surrounding states in $\mathcal{R}$ relative to $\rvec{r}$.

To understand how the MRSE behaves with the Bayesian-region properties in the asymptotic limit of $N$, it is necessary to calculate the MRSE in this limit. After some straightforward calculations in \ref{app:MRSE_exp1}, it turns out that for sufficiently large $N$ where Case~1 holds, the RSE takes the simple form
\begin{eqnarray}
\RSE{}{\rvec{r}}=\norm{\ML{\rvec{r}}-\rvec{r}}^2+2\Tr\{\FML^{-1}\}\dfrac{(-\log\lambda)}{d+2}\,.
\label{eq:MRSE_exp1}
\end{eqnarray}
We observe that the RSE linearly fuses the regular ``frequentist'' point-estimator accuracy measure, the squared error of $\ML{\rvec{r}}$ for a fixed unknown $\rvec{r}$, with ``Bayesian'' elements that characterize the region $\mathcal{R}$. Evidently, we get $\RSE{}{\rvec{r}}=\MSE{\rvec{r}}$ for $\lambda=1$. With this, we may invoke the property $\MEAN{}{\ML{\rvec{r}}}\rightarrow\rvec{r}$ for sufficiently large $N$ and arrive at the formula
\begin{eqnarray}
\mathrm{MRSE}(\rvec{r})=\Tr\{\dyadic{F}^{-1}\}\left(1-\dfrac{2\,\log\lambda}{d+2}\right)
\label{eq:MRSE_exp2}
\end{eqnarray}
for the MRSE, where we have implicitly assumed that $\FML\approx\dyadic{F}=N\dyadic{F}_1$ in the asymptotic limit and $\dyadic{F}_1$ is the Fisher information evaluated with $\rvec{r}$ for a single copy $N=1$ of datum.\footnote{We recall that all data copies are i.i.d., such that $\FML\approx\dyadic{F}$ is an $N$ multiple of $\dyadic{F}_{\textsc{ml},1}\approx\dyadic{F}_1$.} For convenience, we shall drop the parametric variable $\lambda$ hereafter.

\subsection{Duality actions on credible-region accuracy}

Equation~\eqref{eq:MRSE_exp2} provides a basis for us to discuss the effects on the accuracy of credible regions depending on how an observer chooses to optimize the region qualities. We first emphasize that the action of fixing the region size while increasing credibility and that of fixing the region credibility while reducing the size are dual actions in the sense that after these actions, the credible region is optimally defined~\cite{Shang:2013cc}. Armed with the concept of region accuracy, we can endow the effects from these dual strategies with richer statistical meaning. To this end, we analyze the uniform-prior MRSE for different parameter dimension $d$ values.

\subsubsection{$d=1$}

In single-parameter estimation, the Fisher information is a numerical quantity $F$ that is related to both the size $s$ and credibility $c$ by
\begin{eqnarray}
F=\dfrac{8}{s^2 V_{\mathcal{R}_0}^2}\Gamma^{-1}_{1/2}(1-c)\,,
\label{eq:Fsc_d1}
\end{eqnarray}
the $d=1$ version of \eqref{eq:sc}. The resulting MRSE for credible intervals is then given by
\begin{eqnarray}
\MRSEc{r;s,c}=\dfrac{s^2 V_{\mathcal{R}_0}^2}{8}\left[\dfrac{1}{\Gamma^{-1}_{1/2}(1-c)}+\dfrac{2}{3}\right]
\label{eq:mrse_d1}
\end{eqnarray}
as a function of $s$ and $c$.

For $d=1$, the influence of the region dual operations on the MRSE is clear. When the credible region size $s=s_0$ is fixed, increasing the credibility $c$ reduces the MRSE for a given $r$, as $\Gamma^{-1}_{d/2}(1-c)$ is a (strictly) monotonically increasing function of $c$ for any $d$. If $c=c_0$ is fixed instead, then reducing $s$ would, of course, reduce the MRSE. Therefore both dual strategies \emph{increases} the region accuracy.

At first sight, $\MRSEc{r;s,c}$ in \eqref{eq:mrse_d1} is apparently independent of $\rvec{r}$. This thought is misleading because as a matter of fact, $s$ and $c$ are related to each other through $F$ as stated in \eqref{eq:Fsc_d1}. If $F$ is allowed to vary by changing the measurement setup or procedure, then $s$ and $c$ would behave as independent variables. Upon reviewing the dual strategies once more, increasing $c$ for a fixed $s=s_0$ or decreasing $s$ for a fixed $c=c_0$ both require an increase in the Fisher information $F$, which is really the underlying physical quantity that controls the mechanisms behind the dual actions. Hence, increasing $c$ for a fixed $s=s_0$ is dual to decreasing $s$ for a fixed $c=c_0$ in the sense that they both reduces the MRSE for credible intervals.

\subsubsection{$d\geq2$}

Here, matters are slightly less straightforward, for the MRSE depends on a more complicated function of the Fisher information $\dyadic{F}$ that is no longer a numerical value. Both $s$ and $c$ are related to each other through $\DET{\dyadic{F}}$, while the MRSE is a function of $\Tr\{\FML^{-1}\}$ and $-\log\lambda$.

We may first consider the case where $s=s_0$. This sets up the constraint
\begin{eqnarray}
\left(\dfrac{s_0 V_{\mathcal{R}_0}}{V_d}\right)^2\DET{\dyadic{F}}=\left[2\,\Gamma^{-1}_{d/2}(1-c)\right]^d
\end{eqnarray}
for $c$ and the functional dependence of MRSE on $\dyadic{F}$ is now elucidated. To increase $c$ (reduce $\lambda$) under a fixed $s=s_0$, it is clear that $\DET{\dyadic{F}}$ should increase so that $\lambda$ decreases in order to maintain a fixed size. However, since $\Tr\{\dyadic{F}^{-1}\}$ is \emph{not} a function of $\DET{\dyadic{F}}$, there is generally no guarantee that the MRSE will decrease with increasing $\DET{\dyadic{F}}$. There is however a trend that this is the case, and this statement can be made more precisely by considering the largest $\Tr\{\dyadic{F}^{-1}\}$ for a given $\DET{\dyadic{F}}$. If we make use of the fact that for any given physical system, the Fisher information must be trace class~\cite{Nordebo:2012fi} ($\Tr\{\dyadic{F}\}\leq B$ for some positive constant $B$), then one can derive the simple inequality 
\begin{eqnarray}
\Tr\{\dyadic{F}^{-1}\}\leq\dfrac{d\,B^{d-1}}{\DET{\dyadic{F}}}
\label{eq:trFinv}
\end{eqnarray}
for a given $\DET{\dyadic{F}}$ (Refer to \ref{app:trFinv} for a short derivation). The stated upper bound is loose for $d\geq2$, but is sufficient to make our case. After invoking the constraint, we then have 
\begin{eqnarray}
\MRSEc{r;s_0,c}\leq&\,\left(\dfrac{s_0 V_{\mathcal{R}_0}}{V_d}\right)^2\dfrac{d\,B^{d-1}}{\left[2\,\Gamma^{-1}_{d/2}(1-c)\right]^d}\left[1+\dfrac{2\,\Gamma^{-1}_{d/2}(1-c)}{d+2}\right]\,.
\end{eqnarray}
We see that for $d=1$, the upper bound above reduces to the exact expression in \eqref{eq:mrse_d1}. Otherwise, this upper bound decreases monotonically with increasing $c$ for $d\geq2$. The same arguments apply when $c=c_0$, only that now $\MRSEc{\rvec{r};s,c_0}\leq\text{const.}\times s^2$ and so decreasing $s$ by increasing $\DET{\dyadic{F}}$ [see \eqref{eq:sc}] reduces the upper bound quadratically.

\subsection{Duality actions on the plausible-region accuracy}
\label{subsec:plausacc}

If we take $\lambda=\lambda_\text{crit}$, then this time, the dual strategies are carried out with the additional constraint imposed on the value of $\lambda$. Hence, $s$ and $c$ are no longer independent variables. Nonetheless, we may still choose to reduce $s$ or increase $c$ subject to this plausible-region condition.

\subsubsection{$d=1$}

For single-parameter estimation, if we choose to increase $c$, then since 
\begin{eqnarray}
-\log\left(2\pi F^{-1}/V_{\mathcal{R}_0}^2\right)=2\,\Gamma^{-1}_{1/2}(1-c)
\end{eqnarray}
we encounter the simple formula
\begin{eqnarray}
\MRSEp{r;c}=\dfrac{V_{\mathcal{R}_0}^2}{2\pi}\,\E{-2\,\Gamma^{-1}_{1/2}(1-c)}\left[1+\dfrac{2\,\Gamma^{-1}_{1/2}(1-c)}{3}\right]\,.
\end{eqnarray}
It follows that increasing $c\geq1-\Gamma(1/2,1/2)/\sqrt{\pi}$ through increasing $F$ appropriately, reduces $\MRSEp{r,c}$ monotonically.

On the other hand, if we choose to reduce $s$, then based on the one-dimensional identity 
\begin{eqnarray}
s=\dfrac{2}{V_{\mathcal{R}_0}\sqrt{F}}\left[\log\left(\dfrac{V_{\mathcal{R}_0}^2F}{2\pi}\right)\right]^{1/2}
\end{eqnarray}
as well as the parametric form
\begin{eqnarray}
\MRSEp{r;F,s(F)}=\dfrac{1}{F}+\dfrac{1}{12}\left(s V_{\mathcal{R}_0}\right)^2\,,
\end{eqnarray}
it turns out that the way to do this is, again, to increase $F\geq2\pi\E{}/V_{\mathcal{R}_0}^2$, so that $\MRSEp{r;F,s(F)}$ decreases monotonically.

\subsubsection{$d\geq2$}

Likewise, we may carry out the same analysis for $d\geq2$ by first remembering that raising $\DET{\dyadic{F}}$ does not guarantee a reduction in $\Tr\{\dyadic{F}^{-1}\}$. Therefore using the inequality in \eqref{eq:trFinv}, we can instead look at the upper bound for the MRSE and find that
\begin{eqnarray}
\MRSEp{\rvec{r};c}\leq\dfrac{d\,B^{d-1}V_{\mathcal{R}_0}^2}{(2\pi)^d}\,\E{-2\,\Gamma^{-1}_{d/2}(1-c)}\left[1+\dfrac{2\,\Gamma^{-1}_{d/2}(1-c)}{d+2}\right]\,.
\end{eqnarray}
It can then be shown that if one increases $c\geq1-\Gamma(d/2,1/2)/(d/2-1)!$ by raising $\DET{\dyadic{F}}$, $\MRSEp{\rvec{r};c}$ decreases monotonically.

The same goes for the strategy of reducing $s$ by increasing $\DET{\dyadic{F}}\geq(2\pi\E{})^d/V_{\mathcal{R}_0}^2$, the upper bound of the parametric expression
\begin{eqnarray}
\MRSEp{\rvec{r};\dyadic{F},s(\dyadic{F})}\leq&\,\dfrac{d\,B^{d-1}}{\DET{\dyadic{F}}}\left[1+\left(\dfrac{s V_{\mathcal{R}_0}}{V_d}\right)^{2/d}\dfrac{\DET{\dyadic{F}}^{1/d}}{d+2}\right]
\end{eqnarray}
decreases monotonically.

\subsection{The short summary}

We can now draw some succinct yet important conclusions, for any trace-class Fisher information $\dyadic{F}$, regarding the statistical meaning of the dual strategies with Bayesian regions of uniform priors. For credible regions, the action of increasing $c$ with a fixed $s$ and its dual action either reduces the MRSE when $d=1$, or its upper bound (set by a physical upper limit of $\Tr\{\dyadic{F}\}$) when $d\geq2$. The remarks for plausible regions are highly similar. Under the constraint $\lambda=\lambda_{\rm{crit}}$, if an observer either increases $c$ or decreases $s$ for the range $\lambda_{\rm{crit}}\leq\E{-1/2}\approx0.6065$, then either the MRSE ($d=1$) or its upper bound ($d\geq2$) drops monotonically (see \ref{app:threshold} for the derivation of plausible-region threshold values for which these behaviors hold).

So the dual operations for credible regions, or their constrained versions for plausible regions, precisely enhance the region accuracy for $d=1$, or produce the tendency to do so for $d\geq2$. As a final note, the upper bound in \eqref{eq:trFinv} used to argue the general tendency in reducing the MRSE with the dual strategies for $d\geq2$ may be tightened if so desired. The conclusions are then further strengthened with these tighter bounds.

\subsection{Situations for Case 2 and 3 from~\cite{BR:part1}}
\label{subsec:cases23}

A general MRSE formula that rigorously accounts for the occurrences of  Case 2 and 3 is difficult to compute, and there is no known exact relations with the duality actions when these cases are incorporated. However under the condition of large $N$, we may state, with proof, the following conservativeness property for categorically assuming only Case 1 in calculating the MRSE: 
\goodline
\emph{{\bf Conservativeness property}---For the primitive prior, if we assume that $N$ is large enough, so that the region boundary $\partial\mathcal{R}\cap\partial\mathcal{R}_0$ is almost flat (refer to~\cite{BR:part1} for the relevant arguments) and statistical fluctuation is small enough such that $\rvec{r}\approx\ML{\rvec{r}}$, then calculating the data average of \eqref{eq:MRSE_exp1} [approximated with \eqref{eq:MRSE_exp2}] always produces a larger value than the actual MRSE for any $d$.}\vspace{-2ex}
\goodline
The main outline of the proof is to show that since the MRSE is the data average of the RSE, if we categorically insist that Case 1 happens when in fact Case 2 or 3 has actually happened, then the corresponding \emph{as-if} RSE is always larger than the actual RSE under the large data-sample condition. Indeed, this categorical RSE is precisely given by \eqref{eq:MRSE_exp1} evaluated with $\ML{\rvec{r}}\in\mathcal{R}$\footnote{Even when Case 3 does happen but Case 1 is assumed, under the large-$N$ limit, the expression in Eq.~(13) of \cite{BR:part1} approaches the actual likelihood that peaks at the correct unrestricted maximum, so that the calculations in \ref{app:MRSE_exp1} also asymptotically yield the categorical RSE.}. Once this is settled, the resulting MRSE estimate, which is \emph{approximated} by \eqref{eq:MRSE_exp2}, is in principle an overestimate.

The conservativeness of averaging \eqref{eq:MRSE_exp1} is clear for $d=1$. The categorical RSE reads 
\begin{eqnarray}
\mathrm{RSE}_\textsc{cat}=\mathrm{RSE}_{d=1}(r;a,\ML{r})=\frac{1}{3} \left[a^2-2\,\ML{r}(a+3 r)+3 r^2+4\,\ML{r}^2\right]\,,
\end{eqnarray}
which is the RSE for the \emph{as-if} Bayesian interval $[a,2\,\ML{r}-a]$ centered at $\ML{r}$. Here $a$ is data dependent. On the other hand, the RSE of the actual Bayesian interval $[a,b]$ is
\begin{eqnarray}
\mathrm{RSE}=\mathrm{RSE}_{d=1}(r;a,b,\ML{r})=\frac{1}{3}\left(a^2+a b+b^2\right)-r(a+b)+r^2\,,
\end{eqnarray}
where $b\in\partial\mathcal{R}_0$ satisfies $a<\ML{r}\leq b\leq 2\,\ML{r}-a$, and the true parameter $a\leq r\leq b$. Consistently, we have $\mathrm{RSE}=\mathrm{RSE}_\textsc{cat}$ when $b=2\,\ML{r}-a$. It is clear that if $r\approx\ML{r}$, 
\begin{eqnarray}
\mathrm{RSE}-\mathrm{RSE}_\textsc{cat}\approx\frac{1}{3}(b-r)(a+b-2 r)\leq0\,,
\end{eqnarray}
which also means that the categorical RSE is an overestimate of the actual RSE.

One can also find an example for which this is not the case, more so if $r$ is far from $\ML{r}$ or $N$ is not large enough. We may inspect the difference
\begin{eqnarray}
\mathrm{RSE}-\mathrm{RSE}_\textsc{cat}=\dfrac{1}{3}(a+b-2\,\ML{r}) (b-3r+2\,\ML{r})
\end{eqnarray}
and find that in order for it to be positive, we simply need $\ML{r}<(3r-b)/2$. This shows that when $d=1$, $\mathrm{RSE}_\textsc{cat}\geq\mathrm{RSE}$ only if $r\approx\ML{r}$, which asymptotically holds in the large-$N$ limit. Armed with the insights from $d=1$, \ref{app:lemma} separately proves the conservativeness property for arbitrary $d$.  

\section{Adaptive methods for optimizing region accuracy}
\label{sec:adap}

\subsection{Optimization of region accuracy}

\begin{figure}[t]
	\centering
	\includegraphics[width=1\columnwidth]{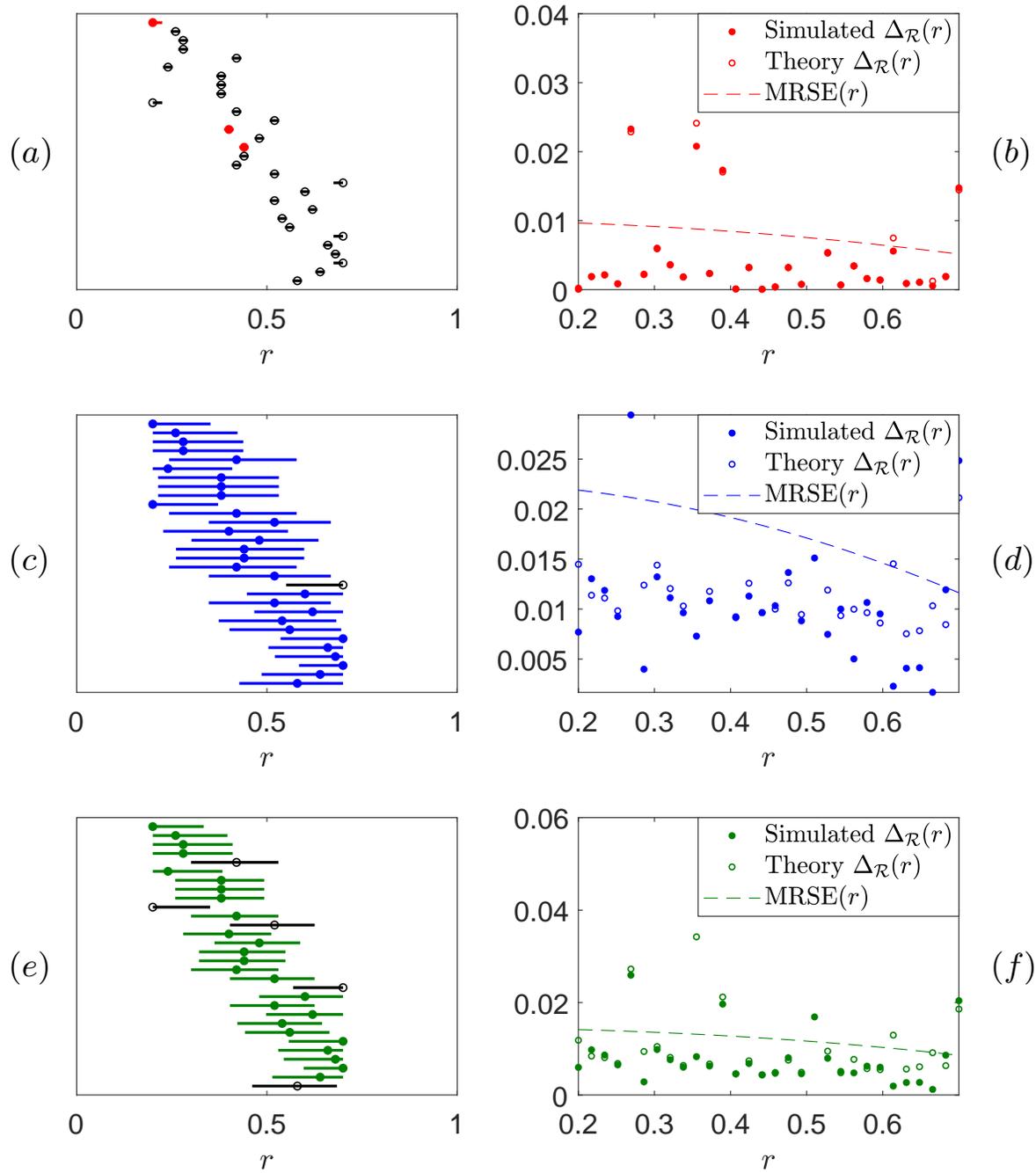}
	\caption{\label{fig:int_d1_1}Plots of (a,c) uniform-prior credible and (e) plausible intervals, as well as (b,d,f) the corresponding RSE and MRSE quantities for a uniform sample of $r$ constrained within the $\mathcal{R}_0$ interval $0.2\leq r\leq0.7$ and a fixed two-outcome POM [$p_1=(1+r)/2,p_2=1-p_1$] used for collecting data of sample size $N=100$. Circular markers denote the ML estimators, and lines with filled circles mark intervals that contain $r$ whereas those with empty circles mark intervals that do not, with the correct probability dictated by $c$ as it should be. Panels (a) and (b) depict the situation of a fixed $s=s_0=0.05$, and panels (c) and (d) concern that of a fixed $c=c_0=0.95$. Every interval in (a) and (c) are constructed with a single dataset. The theoretical RSE [from \eqref{eq:MRSE_exp1}] does not usually match the simulated one since $N$ is small.}
\end{figure}

\begin{figure}[t]
	\centering
	\includegraphics[width=1\columnwidth]{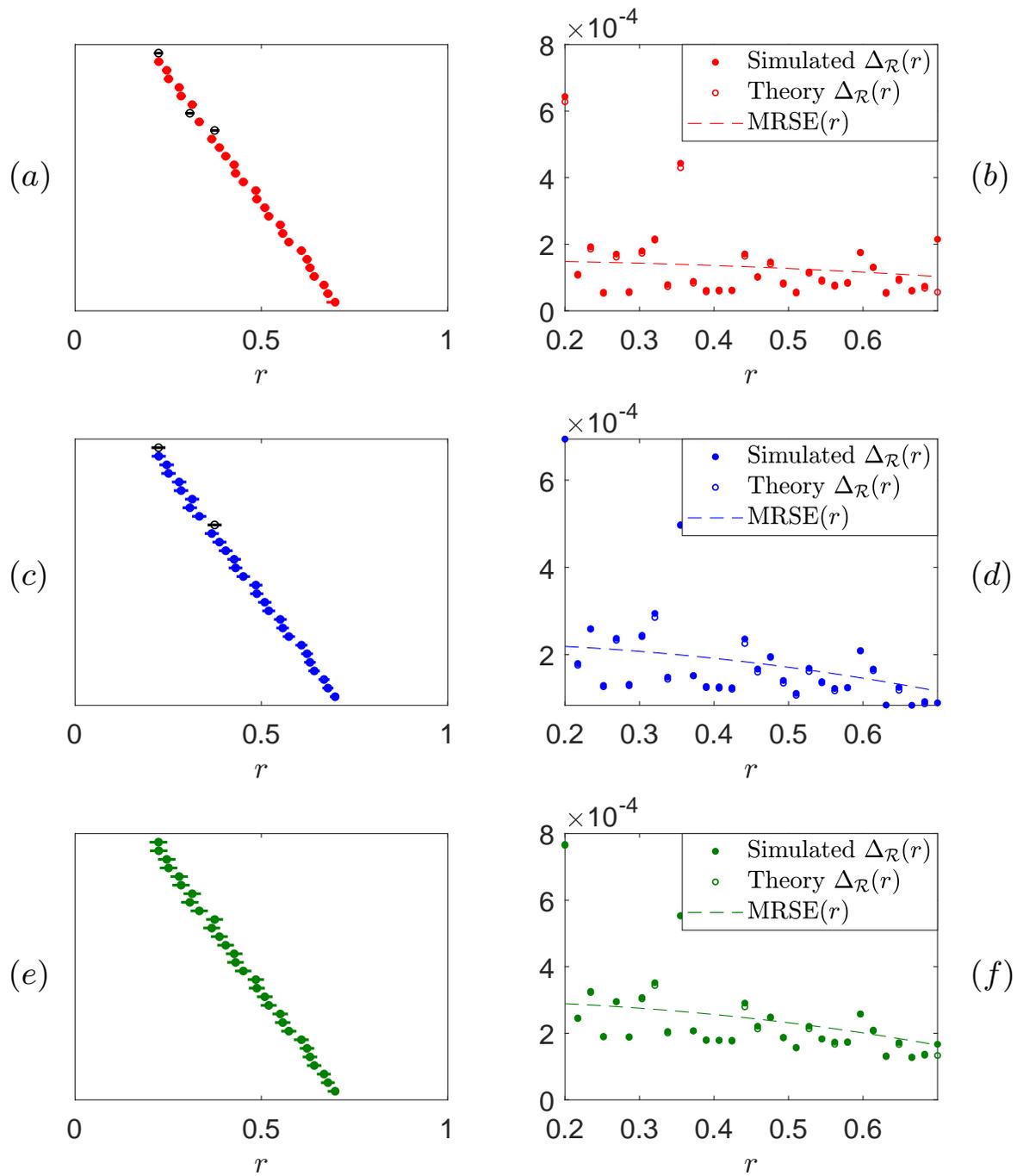}
	\caption{\label{fig:int_d1_2}Similar plots as in Fig.~\ref{fig:int_d1_1} for $N=10000$, where all other specifications remain unchanged. The RSE and MRSE are on average lower than those in Fig.~\ref{fig:int_d1_1} by about two orders of magnitude. This is consistent with the way the credible intervals respond to an increase in $N$. When $c=0.95$, all intervals become shorter, and when $s=0.05$, the intervals adjust their centers to increase the statistical coverage of $r$. All intervals eventually align to minimize the average distance from $r$.\\[0.5cm]}
\end{figure}

Our next goal is to device methods that minimizes the MRSE for any $\rvec{r}$ as defined in \eqref{eq:MRSE_exp2}. For $d=1$, maximizing the determinant of the Fisher information directly reduces the MRSE according to the assessments in Sec.~\ref{sec:statacc}, as the determinant is simply the numerical Fisher information itself. To optimize $F=NF_1$ or the MRSE, an observer may choose to either increase $N$ for a fixed POM that defines $F_1$, or optimize $F_1$ over feasible POMs for a given $N$. As an example, Figs.~\ref{fig:int_d1_1} and \ref{fig:int_d1_2} express what happens to the MRSE when $N$ is increased when a fixed two-outcome POM is used to perform single-parameter estimation. In what follows, we shall address the more interesting problem of optimizing the MRSE, and its optimization for $d=1$ is equivalent to the search for the optimal POM that approaches the well-known quantum Fisher information $F\leq F_\text{Q}$~\cite{Yuen:1973aa,Helstrom:1974aa,Braunstein:1994sd} subject to either a fixed $s$ or $c$ when reporting credible regions, or $\lambda=\lambda_{\rm{crit}}$ when reporting plausible regions. 

When $d\geq2$, the lesson learnt from Sec.~\ref{sec:statacc} shows that the maximization of $\DET{\dyadic{F}}$ does \emph{not} guarantee a minimization of the MRSE. In spite of this, an observer may still carry out POM optimization to minimize the MRSE subject to the kind of Bayesian region that he or she is interested in reporting along with the ML estimator. This is essentially the region analog of maximizing the quantum Fisher information by virtue of Eq.~\eqref{eq:MRSE_exp2}. The correct maximum depends on the true parameter $\rvec{r}$, which is always unknown to the observer. In view of this, we shall develop numerical adaptive protocols that require only the measured data and $d$ to carry out the MRSE minimization. These protocols are applicable in practical experimental situations where the measurement settings for the POM are described by the variable $\rvec{m}\,\widehat{=}\,\TP{(m_1\,\,m_2\,\,\ldots\,\,m_{d_m})}\in\mathcal{S}_{\rvec{m}}$ with a finite dimension $d_m$.

\subsection{Adaptive scheme for credible regions}
\label{subsec:cred_sch}

Without knowing $\rvec{r}$, the experimental data $\mathbb{D}$, if IC in the $d$-dimensional vector space, can still give us a unique ML estimator $\ML{\rvec{r}}$. In the limit of large $N$, this asymptotically unbiased ML estimator is also presumed to be statistically consistent ($\ML{\rvec{r}}\rightarrow\rvec{r}$). With these good properties, $\ML{\rvec{r}}$ can be used as the \emph{a posteriori} state in place of $\rvec{r}$, with which we can make educated guesses for the optimal settings that minimize the MRSE, where its asymptotic expressions in terms of $\dyadic{F}$ are given as
\begin{eqnarray}
\MRSEc{\dyadic{F};s=s_0}&=&\,\Tr\{\dyadic{F}^{-1}\}\left[1+\left(\dfrac{s_0 V_{\mathcal{R}_0}}{V_d}\right)^{2/d}\dfrac{\DET{\dyadic{F}}^{1/d}}{d+2}\right]\,,\nonumber\\
\MRSEc{\dyadic{F};c=c_0}&=&\,\Tr\{\dyadic{F}^{-1}\}\left[1+\dfrac{2\Gamma^{-1}_{d/2}(1-c_0)}{d+2}\right]\,.
\end{eqnarray} 

As a related side note on numerically implementing the adaptive schemes, we mention that when $N$ or $d$ is too large for Monte~Carlo numerical calculations of $s$ and $c$ to work, the asymptotic tools presented in \cite{BR:part1} may be used.

Since $s$ and $c$ are independent, we have the following adaptive algorithm that carries out a total of $K$ adaptive steps for a total of $N$ measurement copies ($N/K$ copies measured in each step) and fixed region size starting with Step~$k=1$:

\noindent
\begin{minipage}[c][9.2cm][c]{1\columnwidth}
\noindent
\rule{\columnwidth}{0.4pt}\\[-2.5ex]
\rule{\columnwidth}{0.4pt}\\
\textbf{MRSE minimization for credible regions}
\begin{enumerate}
	\item\label{sstep:data} Collect $\mathbb{D}_k$ with the setting $\rvec{m}_k$ and compute the $d$-dimensional $\ML{\rvec{r}}$ with the accumulated dataset $\{\mathbb{D}_1,\mathbb{D}_2,\ldots,\mathbb{D}_k\}$.
	\item\label{sstep:c} Set the \emph{a posteriori} state $\rho_0=\ML{\rvec{r}}$, and generate $L$ simulated datasets from $\rho_0$ for each of $n_m$ measurement variables. Here $n_m$ should be a reasonable number of measurement-setting variables $\{\rvec{m}_j\}^{n_m}_{j=1}$ that uniformly covers $\mathcal{S}_{\rvec{m}}$.
	\item With a total of $L n_m$ simulated and $k$ measured datasets, obtain the projected ML estimators $\left\{\widetilde{\ML{\rvec{r}}}_{j,l}\right\}^{n_m,L}_{j=1,l=1}$ and the corresponding projected ML Fisher information $\left\{\widetilde{\FML}_{j,l}\right\}^{n_m,L}_{j=1,l=1}$.
	\item Minimize the MRSE [evaluated with the projected Fisher information using and averaged over all $L$ datasets] with \emph{either} $s=s_0$ \emph{or} $c=c_0$ and set to measure the optimal $\rvec{m}_{\text{opt},k}$.
	\item Increase $k$ by one and repeat Steps~(\ref{cpstep:data}) through (\ref{cpstep:choose}) until $k=K$.	\\[-5ex]
\end{enumerate}
\rule{\columnwidth}{0.4pt}\\[-2.5ex]
\rule{\columnwidth}{0.4pt}
\end{minipage}

\subsection{Adaptive scheme for plausible regions}
\label{subsec:plaus_sch}

A similar adaptive protocol can be developed to minimize
\begin{eqnarray}
\MRSEp{\dyadic{F}}=\Tr\{\dyadic{F}^{-1}\}\left[1+\dfrac{\log\left(V_{\mathcal{R}_0}^2/(2\pi)^d\right)}{d+2}+\dfrac{\log\left(\DET{\dyadic{F}}\right)}{d+2}\right]\,:
\end{eqnarray}

\noindent
\begin{minipage}[c][9.2cm][c]{1\columnwidth}	
	\noindent
	\rule{\columnwidth}{0.4pt}\\[-2.5ex]
	\rule{\columnwidth}{0.4pt}\\
	\textbf{MRSE minimization for plausible regions}
	\begin{enumerate}
		\item\label{cpstep:data} Collect $\mathbb{D}_k$ with the setting $\rvec{m}_k$ and compute the $d$-dimensional $\ML{\rvec{r}}$ with the accumulated dataset $\{\mathbb{D}_1,\mathbb{D}_2,\ldots,\mathbb{D}_k\}$.
		\item\label{cptep:c} Set the \emph{a posteriori} state $\rho_0=\ML{\rvec{r}}$, and generate $L$ simulated datasets from $\rho_0$ for each of $n_m$ measurement variables. Here $n_m$ should be a reasonable number of measurement-setting variables $\{\rvec{m}_j\}^{n_m}_{j=1}$ that uniformly covers $\mathcal{S}_{\rvec{m}}$..
		\item With a total of $L n_m$ simulated and $k$ measured datasets, obtain the projected ML estimators $\left\{\widetilde{\ML{\rvec{r}}}_{j,l}\right\}^{n_m,L}_{j=1,l=1}$ and the corresponding projected ML Fisher information $\left\{\widetilde{\FML}_{j,l}\right\}^{n_m,L}_{j=1,l=1}$.
		\item\label{cpstep:choose} Minimize the MRSE [evaluated with the projected Fisher information and averaged over all $L$ datasets] with $\lambda=\lambda_{\rm{crit}}$ and set to measure the optimal $\rvec{m}_{\text{opt},k}$.
		\item Increase $k$ by one and repeat Steps~(\ref{cpstep:data}) through (\ref{cpstep:choose}) until $k=K$.	\\[-5ex]
	\end{enumerate}
	\rule{\columnwidth}{0.4pt}\\[-2.5ex]
	\rule{\columnwidth}{0.4pt}
\end{minipage}

\subsection{Differences from known Bayesian adaptive schemes}

Before we proceed with the examples, it is timely to mention here that there exist adaptive schemes that choose optimal configurations for enhancing the tomographic quality of \emph{point estimators}, for instance, in tracking drifts in quantum states and processes~\cite{Granade:2012aa,Granade:2016aa}. The primary mechanism behind these adaptive schemes is to improve accuracies of point estimators measured by objective functions of the posterior Hessian that encodes geometrical properties of Bayesian region around the posterior maximum in the limit of large $N$.

We emphasize that these previously proposed schemes are of a different qualitative nature from that of the adaptive MRSE minimization schemes presented here. The present concern is the accuracy of an error region, as opposed to a single estimator. In this case, not only are the geometrical properties of the Bayesian region $\mathcal{R}$ around $\ML{\rvec{r}}$ important in our considerations, but also the quality of every state within $\mathcal{R}$ relative to the unknown true parameter $\rvec{r}$. Maximizing the MRSE therefore operates on a higher hierarchical level---it is the whole error region $\mathcal{R}$, namely the point estimator $\ML{\rvec{r}}$ and surrounding error states, that collectively possesses the maximum (average) accuracy (minimum MRSE), not just $\ML{\rvec{r}}$. 

That being said, the idea of region accuracy and its maximization not only forms one bridge that connects parts of frequentist and Bayesian elements, but also directly support the Bayesian spirit that surrounding states of $\ML{\rvec{r}}$ are just as important (according to the prior) in parameter error analysis. The adaptive methods established in Sec.~\ref{subsec:cred_sch} and \ref{subsec:plaus_sch} are meant for this distinct purpose.

\begin{figure}[b]
	\centering
	\includegraphics[width=1\columnwidth]{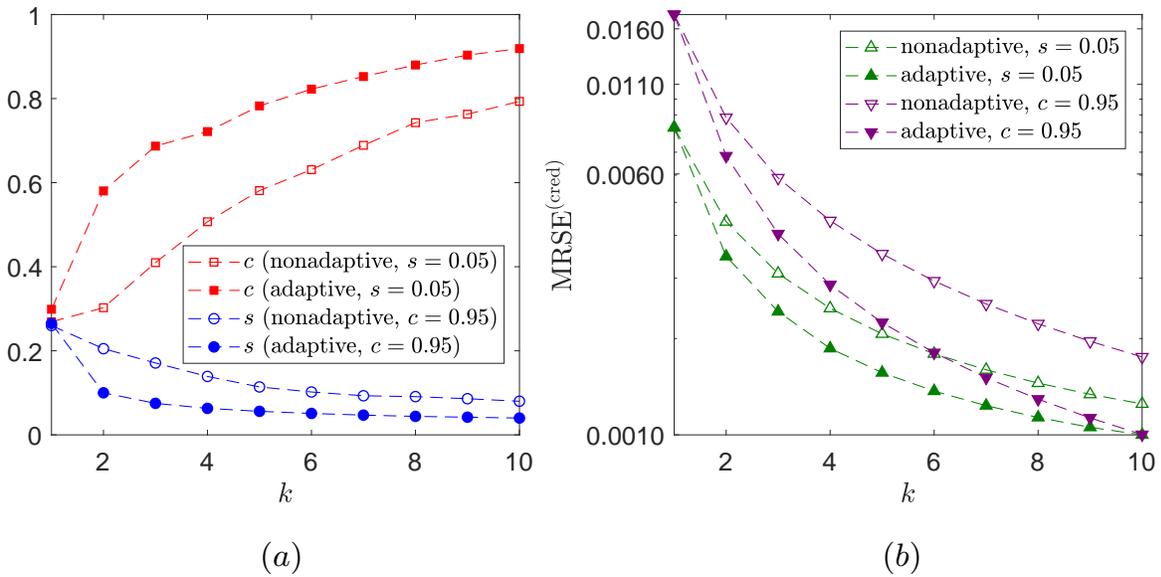}
	\caption{\label{fig:phase_cred}
		Plots of (a) the credible-interval properties and (b) $\mathrm{MRSE}^{\text{(cred)}}$ for $r=1.179$ and a fixed squeeze parameter $\zeta=0.7$. While all numerical schemes are executed with only ML estimators, the MRSE graphs are plotted with the true parameter to show the correct accuracies, just like any analysis of the point MSE. A primitive prior that extends to the finite range $0\leq\phi\leq\pi/2$ is assumed in the simulation as prior knowledge about the unknown relative phase $r$. Here $N=1000$ copies are distributed equally to $K=10$ adaptive steps that are carried out by each adaptive protocol. The nonadaptive versions measure the fixed LO phase $\vartheta=1.837$ throughout the run, which is less efficient than their adaptive counterparts that begin with the same LO phase and eventually converge to the optimal LO phase subject to the constraint imposed on $\mathcal{R}$ (fixed $s$ or $c$).
	}
\end{figure}

\begin{figure}[t]
	\centering
	\includegraphics[width=1\columnwidth]{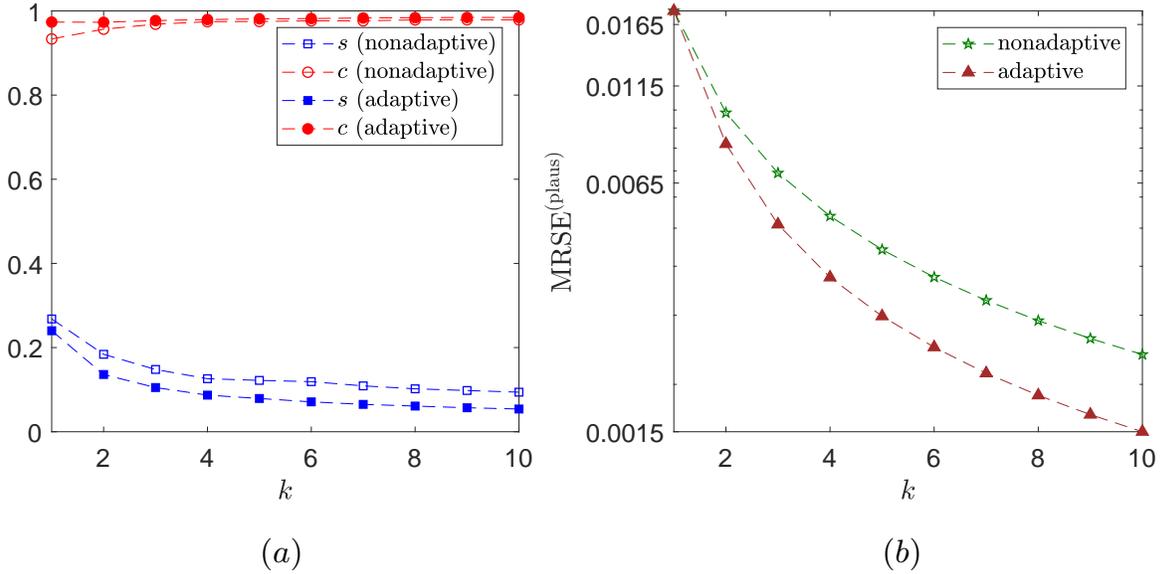}
	\caption{\label{fig:phase_plaus}Plots of (a) the plausible-interval properties and (b) $\mathrm{MRSE}^{\text{(plaus)}}$ for $\phi=1.179$ with identical specifications as in Fig.~\ref{fig:phase_cred}. We see a reduction in MRSE when the adaptive scheme is applied for such intervals.}
\end{figure}

\section{Examples}
\label{sec:examples}

\subsection{Phase-shifted homodyne interferometer ($d=1$)}
\label{subsec:ph1D}

An important single-parameter estimation task in quantum information is phase reconstruction for an interferometer with quantum input resources~\cite{Caves:1981aa,Dorner:2009aa,Paesani:2009aa}. A very common type of interferometer is the homodyne measurement setup~\cite{Yuen:1983ba,Abbas:1983ak,Vogel:1989zr,Banaszek:1997ot} that is employed in continuous-variable quantum tomography and cryptography. An interesting case arises when both the source (mode $a$) and local-oscillator (LO) arms of the homodyne setup differ by an unknown relative phase $r=\phi$ that can be modeled by the phase-shifter described by a unitary operator $U(\phi)=\E{\I a^\dag a\,\phi}$. The job is to characterize the unknown phase $\phi$ for the interferometer setup, which is a one-dimensional problem ($d=1$).

It is known in~\cite{Olivares:2009hi} that using a squeezed-vacuum state $\ket{\zeta}\bra{\zeta}$ for mode $a$ saturates the quantum Cram{\'e}r--Rao bound in $\phi$ estimation, where the Born probabilities
\begin{eqnarray}
p(x_\vartheta,\vartheta)&=&\,\dfrac{\E{-x^2_\vartheta/\sigma_{\vartheta,\phi}^2}}{\sqrt{2\pi\sigma_{\vartheta,\phi}^2}}\,,\nonumber\\
\qquad\sigma_{\vartheta,\phi}^2&=&\,\dfrac{1}{2}\left[\cosh(2\zeta)+\cos(2\vartheta-2\phi)\sinh(2\zeta)\right]\,,
\end{eqnarray}
encode the unknown phase $\phi$, the (real) squeeze parameter $\zeta$ and the homodyne local oscillator (LO) phase $\vartheta$.

The adaptive schemes in Sec.~\ref{sec:adap} are readily applicable to this one-dimensional quantum estimation scenario. They equivalently maximize the Fisher information
\begin{eqnarray}
F=N\dfrac{\sinh(2\zeta)^2-\left[\cosh(2\zeta)-2\sigma_{\vartheta,\phi}^2\right]^2}{2\sigma_{\vartheta,\phi}^2}
\end{eqnarray}
for this problem. The optimal LO phase $m_\text{opt}=\vartheta_\text{opt}=\phi-\cos ^{-1}(\tanh\zeta)/2$ that achieves the maximum depends on $\phi$, and the adaptive schemes asymptotically select this value without this knowledge. Figures~\ref{fig:phase_cred} and \ref{fig:phase_plaus} demonstrate the advantage of employing adaptive schemes over nonadaptive ones in increasing region accuracies with a fixed total number of copies $N$ and $\zeta$ by performing the relevant optimization over the space $\mathcal{S}_m$ of LO phase $m=\vartheta$. Figures~\ref{fig:phase_cred} and \ref{fig:phase_plaus} compare the difference in MRSE between adaptive and nonadaptive IC schemes. A more sophisticated second example in quantum metrology shall 
follow.

\subsection{Three-path interferometer ($d=2$)}

One may generalize a typical two-arm interferometer, such as the homodyne setup discussed previously, to a three-arm interferometer (modes $a$, $b$ and $c$) of unknown relative phases $\rvec{r}\,\widehat{=}\,\TP{(\phi_1\,\,\phi_2)}$ in the three arms, with $\phi_1$ being the phase difference between modes $a$ and $c$, and $\phi_2$ between $b$ and $c$. Such an interferometer poses a two-parameter estimation problem and may be modeled with the ordered sequence of first a \emph{beam tritter} ($U_3$), followed by a three-arm phase shifter [$U(\phi_1,\phi_2)$], and finally another beam tritter---$U_\text{three-path}(\phi_1,\phi_2)=U_3U(\phi_1,\phi_2)U_3$---, where
\begin{eqnarray}
U(\phi_1,\phi_2)&=&\,\E{\I (a^\dag a\,\phi_1+b^\dag b\,\phi_2)}\,,\nonumber\\
\qquad\quad U_3&\,\,\widehat{=}&\dfrac{1}{\sqrt{3}}\begin{pmatrix}
1 & \E{2\pi\I/3} & \E{2\pi\I/3}\\
\E{2\pi\I/3} & 1 & \E{2\pi\I/3}\\
\E{2\pi\I/3} & \E{2\pi\I/3} & 1
\end{pmatrix}\,.
\end{eqnarray}

\begin{figure}[t]
	\centering
	\includegraphics[width=1\columnwidth]{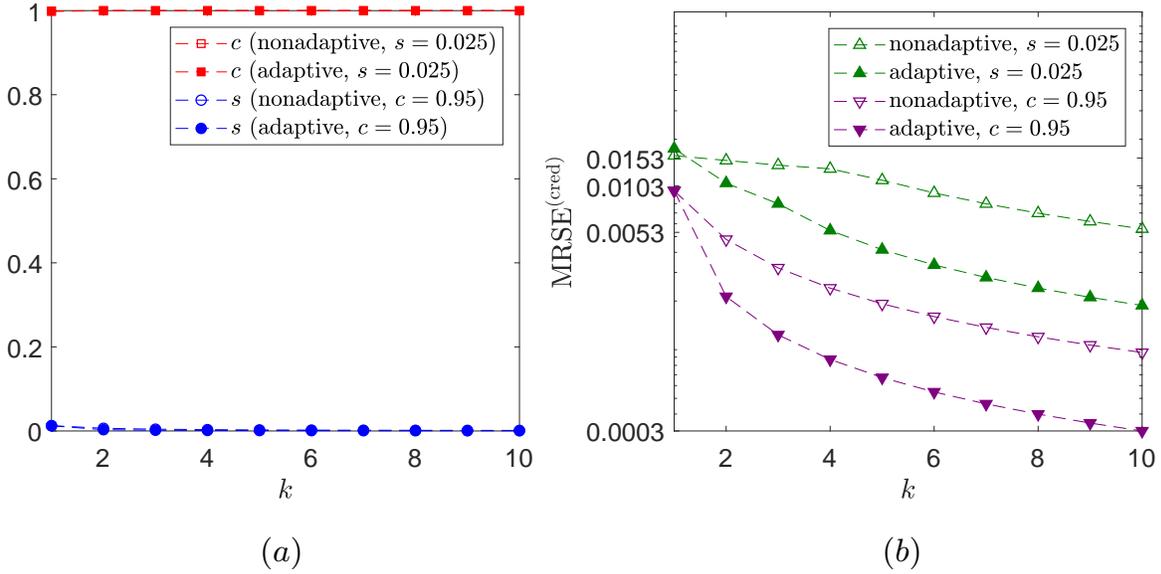}
	\caption{\label{fig:tritter_cred}Plots of (a) the credible-region properties and (b) $\mathrm{MRSE}^{\text{(cred)}}$ for the true parameter pair $(\phi_1,\phi_2)=(0.5,1.0)$ with the step number $k$. In each step, $N=500$ idealized photon-counting events are taken, so that at the end of every run, the observer measures a total of $N=5000$ copies of data. The parameter space $\mathcal{R}_0=\{\phi_1|\,0\leq\phi_1\leq\pi/2\}\times\{\phi_2|\,0\leq\phi_2\leq\pi/2\}$ is defined by the primitive prior with respect to $\phi_1$ and $\phi_2$. Similar to the one-dimensional parameter estimation scenario in Sec.~\ref{subsec:ph1D}, the nonadaptive schemes collect data with a fixed setting $\rvec{m}=(0,0)$, whilst the adaptive schemes actively search for more optimal $\rvec{m}_k$s at every step by analyzing collected data.}
\end{figure}

\begin{figure}[t]
	\centering
	\includegraphics[width=1\columnwidth]{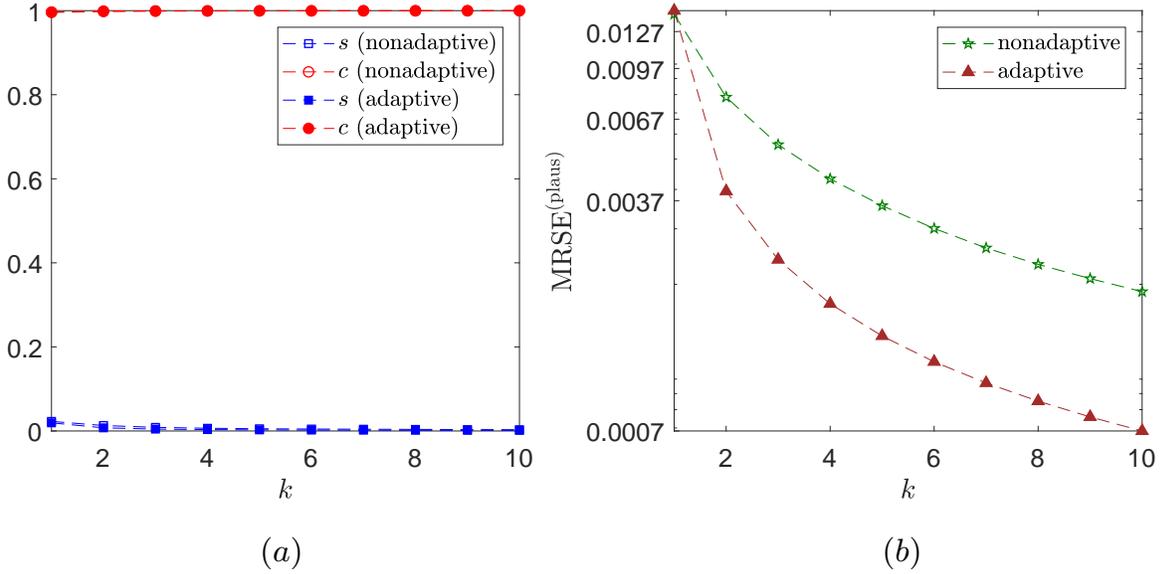}
	\caption{\label{fig:tritter_plaus}Plots of (a) the plausible-region properties and (b) $\mathrm{MRSE}^{\text{(plaus)}}$ for the Gaussian squeezed state of parameters \mbox{$(\phi_1,\phi_2)=(0.5,1.0)$}. Similar behavior is witnessed here with the same figure specifications as Fig.~\ref{fig:tritter_cred}.}
\end{figure}

To estimate a given unknown $\rvec{r}$, we shall suppose that incoming photons are initially in the three-mode input state $\rho=\ket{1,1,1}\bra{1,1,1}$ to be described by a tripartite Fock state, which, after traversing the interferometer that is additionally encoded with measurement control phases $\rvec{m}\,\widehat{=}\,\TP{(\psi_1\,\,\psi_2)}$ for tuning the final estimation accuracy, would then encounter idealized photon-counting detectors that result in the 10 Born probabilities $p_{n_1,n_2,n_3}(\psi_1,\psi_2;\phi_1,\phi_2)=|\opinner{n_1,n_2,n_3}{U_\text{three-path}(\psi_1-\phi_1,\psi_2-\phi_2)}{1,1,1}|^2$ ($n_1+n_2+n_3=3$). We refer the interested Reader to the supplementary information of \cite{Ciampini:2016ma} for detailed calculations of $\dyadic{F}$ and $p_{n_1,n_2,n_3}(\psi_1,\psi_2;\phi_1,\phi_2)$, and instead provide a comparison between adaptive and nonadaptive protocols for such a two-parameter phase estimation problem with Figs.~\ref{fig:tritter_cred} and \ref{fig:tritter_plaus}.

\begin{figure}[t]
	\centering
	\includegraphics[width=1\columnwidth]{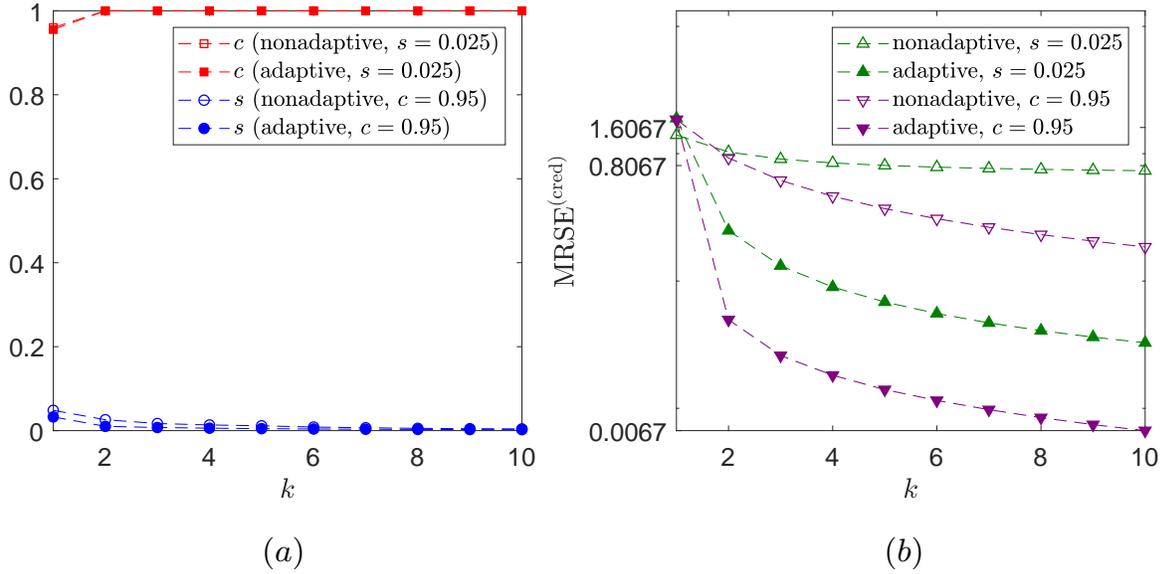}
	\caption{\label{fig:sqst_cred}Plots of (a) the credible-region properties and (b) $\mathrm{MRSE}^{\text{(cred)}}$ for $(\nu,\alpha)=(3.2580,1.0517)$, where $1000$ copies are measured in each step, which tallies to a total of $N=10000$. The primitive prior assigned for the simulations results in the finite parameter space $\mathcal{R}_0=\{\nu|\,1\leq\nu\leq5\}\times\{\alpha|\,0\leq\alpha\leq\pi/2\}$. All schemes, regardless of whether they are adaptive or not, start with the initial LO phase pair \mbox{$(\vartheta_1,\vartheta_2)=(0.27,1.0)$}, and the adaptive schemes find much more optimal phase pairs to achieve the minimum MRSE.}
\end{figure}
\begin{figure}[t]
	\centering
	\includegraphics[width=1\columnwidth]{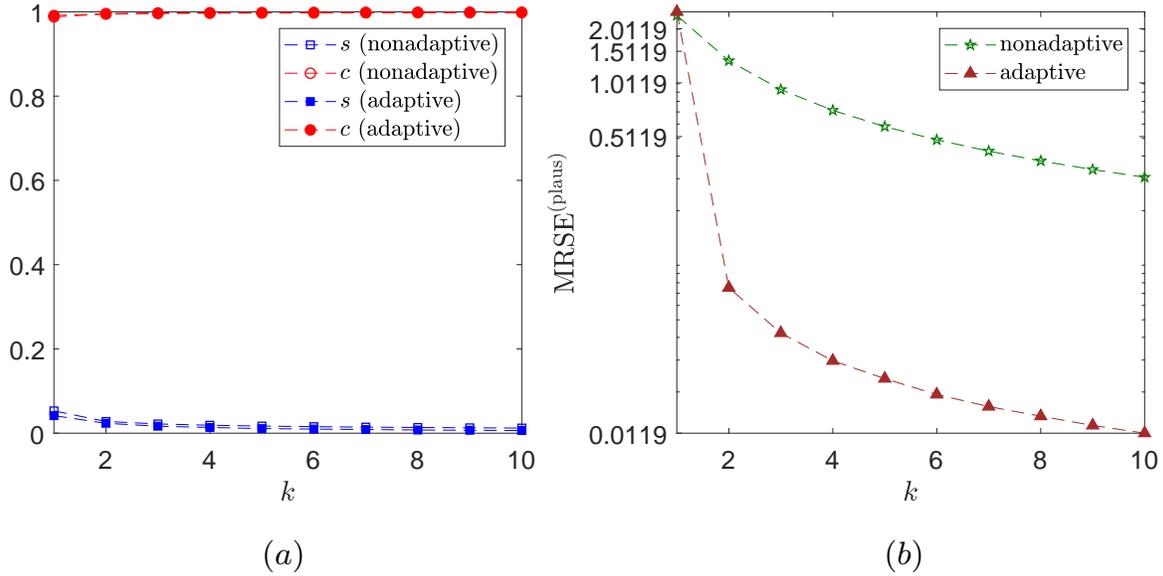}
	\caption{\label{fig:sqst_plaus}Plots of (a) the plausible-region properties and (b) $\mathrm{MRSE}^{\text{(plaus)}}$ for the Gaussian squeezed state of parameters \mbox{$(\nu,\alpha)=(3.2580,1.0517)$}. Once again we see the significant reduction in the MRSE using the adaptive numerical scheme for plausible regions, in contrast with the MRSE corresponding to the nonadaptive one.}
\end{figure}

\subsection{Squeezed-state characterization ($d=2$)}

The third example that we shall investigate is related to Gaussian states, which are important resources in quantum information~\cite{Lorenz:2004aa, Lance:2005aa,Scarani:2009cq,Weedbrook:2012ag}. Every single-mode Gaussian state (of known mean) can be fully specified by the covariance of its Gaussian quasiprobability distribution. For simplicity, we shall again take homodyne detection as the POM for Gaussian-state characterization in this section. For a given orientation angle $\alpha$ of its phase-space quasiprobability distribution with respect to the $x$ phase-space ordinate is known, its temperature $\mu$ and squeeze parameter $\nu$, the covariance of the Gaussian state is given by
\begin{eqnarray}
\dyadic{G}=\dyadic{G}_{\mu,\nu,\alpha}\,\widehat{=}\,\dfrac{\mu}{2\nu}
\begin{pmatrix}
\nu^2(\cos\alpha)^2+(\sin\alpha)^2\,\,&\,\,\left(1-\nu^2\right)\sin (2\alpha)/2\\
\left(1-\nu^2\right)\sin (2\alpha)/2\,\,&\,\,\nu^2(\sin\alpha)^2+(\cos\alpha)^2
\end{pmatrix}\,.
\end{eqnarray}
The task of characterizing $\dyadic{G}$ has been studied in~\cite{Rehacek:2009ys,Rehacek:2015qp,Muller:2016da,Teo:2017aa,Teo:2017uh}. 

An interesting situation is when $\mu$ is preliminarily known (possibly from thermal-light calibrations) and normalized, and we are now interested in characterizing the squeezing properties $\rvec{r}\,\widehat{=}\,\TP{(\nu\,\,\alpha)}$ of this Gaussian state~\cite{Muller:2012ys}. It can be shown that if an IC setting $\rvec{m}\,\widehat{=}\,\TP{(\vartheta_1\,\,\vartheta_2)}$ consisting of a pair of LO phases is measured in such a two-parameter estimation problem, the complete two-dimensional Fisher information $\dyadic{F}=\dyadic{\widetilde{F}}(\vartheta_1)+\dyadic{\widetilde{F}}(\vartheta_2)$ is the sum of its independent Fisher information components, where $\dyadic{\widetilde{F}}(\vartheta)$ contains the elements
\begin{eqnarray}
\widetilde{F}_{11}(\vartheta)&=&\,\frac{\left[\nu ^2+\left(\nu ^2+1\right) \cos (2\alpha+2\vartheta)-1\right]^2}{2\nu^2 \left[\nu ^2 +\left(\nu ^2-1\right)\cos (2\alpha+2\vartheta)+1\right]^2}\,,\nonumber\\
\widetilde{F}_{22}(\vartheta)&=&\,\frac{2 \left(\nu ^2-1\right)^2 [\sin(2\alpha +2\vartheta)]^2}{\left[\nu ^2+\left(\nu ^2-1\right) \cos (2\alpha+2\vartheta)+1\right]^2}\,,\nonumber\\
\widetilde{F}_{12}(\vartheta)&=&\,\frac{\left(1-\nu ^2\right) \sin (2\alpha+2\vartheta) \left[\nu ^2+\left(\nu ^2+1\right) \cos (2\alpha+2\vartheta)-1\right]}{\nu  \left[\nu ^2+\left(\nu ^2-1\right) \cos (2\alpha+2\vartheta)+1\right]^2}\,.
\end{eqnarray}

Figures~\ref{fig:sqst_cred} and \ref{fig:sqst_plaus} illustrate, once again, how adaptive schemes are indeed extremely useful for constructing much more accurate Bayesian regions for ML estimators whenever the observer begins the parameter estimation experiment with poorly chosen measurement setting variables, which frequently occurs as $\rvec{r}$ is unknown.

\section{Conclusion}

The key results of this article revolve around the definition of region accuracy, which is introduced to endow every Bayesian error region with the notion of a frequentist-flavored statistical accuracy (averaged over the entire error region) relative to the unknown true parameter of interest. The region accuracy turns out to do more than just this: it treats the point estimator and its surrounding states within the Bayesian region on equal footing (up to the prior distribution) and endows a mean squared error collectively. This natural concept elucidates the statistical consequences of either minimizing the credible-region size with fixed credibility, or the dual action of maximizing its credibility with fixed size---both actions increase the region accuracy consistent with our intuitive understanding of these Bayesian regions. 

Efforts are then spent on establishing adaptive strategies to optimize region accuracy given only collected data, the dimension of a given estimation problem and no other assumptions about the true parameter. These adaptive procedures are applied to practically interesting examples in quantum metrology and Gaussian-state characterization, all of which agree with their positive estimation performance. We believe that these adaptive numerical schemes, together with the asymptotic techniques in the companion article~\cite{BR:part1}, shall form a useful toolkit for Bayesian-region construction in practical experimental settings where the dimension of the problem and data sample size are at least moderately large.

\section{Acknowledgments}
We are grateful to B.-G.~Englert for a pertinent discussion concerning plausible-region optimization. We acknowledge financial support from the BK21 Plus Program (21A20131111123) funded by the Ministry of Education (MOE, Korea) and National Research Foundation of Korea (NRF), the NRF grant funded by the Korea government (MSIP) (Grant No. 2010-0018295), and the Korea Institute of Science and Technology Institutional Program (Project No. 2E27800-18-P043).

\appendix

\section{The derivation of \eqref{eq:MRSE_exp1}}
\label{app:MRSE_exp1}

Following Appendix~A in \cite{BR:part1}, we write the numerator of $\RSE{}{\rvec{r}}$ as
\begin{eqnarray}
g(\rvec{r})=\left.\int\dfrac{\D t}{2\pi\I}\,\dfrac{\E{-\I t\log\left(\lambda L_\text{max}\right)}}{t-\I\epsilon}\,\int_{\mathcal{R}_0}(\D\,\rvec{r}')\left(\rvec{r}'-\rvec{r}\right)^2\,\E{\I t\log L(\mathbb{D}|\rvec{r}')}\right|_{\epsilon=0}\,,
\end{eqnarray}
where after using the Gaussian approximation for $L(\mathbb{D}|\rvec{r})$
\begin{eqnarray}
&&\,\int_{\mathcal{R}_0}(\D\,\rvec{r}')\left(\rvec{r}'-\rvec{r}\right)^2\,\E{\I t\log L(\mathbb{D}|\rvec{r}')}\nonumber\\
&\approx&\,\dfrac{\E{\I t\log L_\text{max}}}{V_{\mathcal{R}_0}}\int\left(\prod_j\D r''_j\right)\rvec{r}''^2\,\E{-\frac{\I t}{2}\rvec{r}''\bm{\cdot}\FML\bm{\cdot}\rvec{r}''}\nonumber\\
&&+\E{\I t\log L_\text{max}}\dfrac{(\ML{\rvec{r}}-\rvec{r})^2}{V_{\mathcal{R}_0}}\int\left(\prod_j\D r''_j\right)\E{-\frac{\I t}{2}\rvec{r}''\bm{\cdot}\FML\bm{\cdot}\rvec{r}''}.
\label{eq:secmom}
\end{eqnarray}
The first term of \eqref{eq:secmom}, which is the second moment for a multivariate Gaussian distribution, can be calculated by noting the calculus identity $\updelta\DET{\dyadic{A}}/\updelta\dyadic{A}=\DET{\dyadic{A}}\dyadic{A}^{-1}$ for any full-rank $\dyadic{A}$:
\begin{eqnarray}
&&\,\int\left(\prod_j\D r''_j\right)\rvec{r}''^2\,\E{-\frac{\I t}{2}\rvec{r}''\bm{\cdot}\FML\bm{\cdot}\rvec{r}''}\nonumber\\
&=&\,\dfrac{2\I}{t}\Tr\left\{\dfrac{\updelta}{\updelta\FML}\int\left(\prod_j\D r''_j\right)\,\E{-\frac{\I t}{2}\rvec{r}''\bm{\cdot}\FML\bm{\cdot}\rvec{r}''}\right\}\nonumber\\
&=&\,\dfrac{1}{\I t}\left(\dfrac{2\pi}{\I t}\right)^{d/2}\Tr\left\{\DET{\FML}^{-1/2}\FML^{-1}\right\}\,.
\end{eqnarray}
The second term simply amounts to (A.3) in \cite{BR:part1}. Altogether we have
\begin{eqnarray}
g(\rvec{r})&=&\,\dfrac{\DET{\FML}^{-1/2}(2\pi)^{d/2}}{V_{\mathcal{R}_0}}\nonumber\\
&&\times\int\dfrac{\D t}{2\pi\I}\,\dfrac{\E{-\I t\log\lambda }}{t-\I\epsilon}\left[\dfrac{\Tr\{\FML^{-1}\}}{(\I t)^{d/2+1}}+\dfrac{(\ML{\rvec{r}}-\rvec{r})^2}{(\I t)^{d/2}}\right]\,.
\end{eqnarray}
The $t$ integrals can be handled in exactly the same manner depicted in Appendix~A of \cite{BR:part1}, which leads to the final answer.

\section{The derivation of \eqref{eq:trFinv}}
\label{app:trFinv}

For a $d$-dimensional full-rank $\dyadic{F}$ of fixed determinant $\DET{\dyadic{F}}=a$ and trace $\Tr\{\dyadic{F}\}=b$, the largest eigenvalue $\lambda_d$ from the ordered sequence $\lambda_1\leq\lambda_2\leq\ldots\leq\lambda_d$ must satisfy the trivial inequality $\lambda_d\leq b$, and the smallest eigenvalue
\begin{eqnarray}
\lambda_1=\dfrac{a}{\lambda_2\ldots\lambda_d}\geq\dfrac{a}{\lambda^{d-1}_d}\geq\dfrac{a}{b^{d-1}}\,.
\end{eqnarray}
This automatically bounds
\begin{eqnarray}
\Tr\{\dyadic{F}^{-1}\}=\sum^d_{j=1}\dfrac{1}{\lambda_j}\leq\dfrac{d}{\lambda_1}\leq\dfrac{d\,b^{d-1}}{a}
\end{eqnarray}
from above. Then clearly, if $\Tr\{\dyadic{F}\}\leq B$ for some large constant $B$, a property of a trace class Fisher information, the inequality in \eqref{eq:trFinv} is achieved.

As a side remark, we remind the Reader that the occasional $p_j=0$ for some POM and $\rvec{r}$ does not violate the trace-class property of $\dyadic{F}$, since these zero-probability events are ignored when defining $L(\mathbb{D}|\rvec{r})$ in the absence of experimental imperfections.

\section{Threshold values for the dual strategies on plausible regions}
\label{app:threshold}

The task is to decrease the upper bounds of $\MRSEp{\rvec{r};c}$ and $\MRSEp{\rvec{r};\dyadic{F},s(\dyadic{F})}$. For the upper bound of $\MRSEp{\rvec{r};c}$, the relevant function of interest is $y_1(c)=\E{-2\,\Gamma^{-1}_{d/2}(1-c)}\,\Gamma^{-1}_{d/2}(1-c)$, which has one global maximum. This maximum stationary point can be obtain by calculating the first-order derivative
\begin{eqnarray}
\dfrac{\D y_1}{\D c}=\E{-\Gamma^{-1}_{d/2}(1-c)}\left(\dfrac{d}{2}-1\right)!\left[\Gamma^{-1}_{d/2}(1-c)\right]^{1-d/2}\left[1-2\,\Gamma^{-1}_{d/2}(1-c)\right]
\end{eqnarray}
using the derivative identity
\begin{eqnarray}
\dfrac{\D}{\D z}\Gamma^{-1}_{d/2}(z)=-\E{\Gamma^{-1}_{d/2}(z)}\left(\dfrac{d}{2}-1\right)!\left[\Gamma^{-1}_{d/2}(z)\right]^{1-d/2}
\end{eqnarray}
for the inverse incomplete Gamma function. Setting $\D y_1/\D c=0$ then gives the solution $c_\text{max}=1-\Gamma(d/2,1/2)/(d/2-1)!$. To show that $c_\text{max}$ is indeed the maximum, one can calculate the second-order derivative
\begin{eqnarray}
\left(\dfrac{\D}{\D c_\text{max}}\right)^2y_1(c_\text{max})=-2^d\left[(d/2-1)!\right]^2
\end{eqnarray}
evaluated at $c=c_\text{max}$, which is clearly negative. This implies that beyond $c>c_\text{max}$, $y_1(c)$, or the upper bound of $\MRSEp{\rvec{r};c}$, decreases monotonically.

To decrease the upper bound of $\MRSEp{\rvec{r};\dyadic{F},s(\dyadic{F})}$ monotonically, it suffices to obtain the threshold value for $\DET{\dyadic{F}}$ beyond which $s(\dyadic{F})$ drops monotonically. This means we need to look at
\begin{eqnarray}
y_2(x)=\dfrac{1}{\sqrt{x}}\left[\log\left(\dfrac{V_{\mathcal{R}_0}^2 x}{(2\pi)^d}\right)\right]^{d/2}\,,
\end{eqnarray}
which also contains one global maximum. Setting its first-order derivative to zero gives $x_\text{max}=(2\pi\E{})^d/(V_{\mathcal{R}_0}^2)$ and a negative second-order derivative
\begin{eqnarray}
\left(\dfrac{\D}{\D x_\text{max}}\right)^2\,y_2(x_\text{max})=-\left(\dfrac{V_{\mathcal{R}_0}^2}{2\pi}\right)^{5/2}\dfrac{d^{d/2-1}}{2}\,\E{-5d/2}\,,
\end{eqnarray}
as it should be. So if $\DET{\dyadic{F}}$ increases beyond the threshold of $(2\pi\E{})^d/(V_{\mathcal{R}_0}^2)$, then $s$ will decrease monotonically. These two threshold values (one for $s$ and one for $c$) coincide with the common value $\lambda_{\rm{crit}}=\E{-1/2}\approx0.6065$.

\section{The conservativeness of averaging \eqref{eq:MRSE_exp1}}
\label{app:lemma}

By invoking the asymptotic techniques in \cite{BR:part1} used to cope with Case 2 and 3, we recall that the relevant Bayesian region $\mathcal{R}$ is essentially a truncated hyperellipsoid with $\partial\mathcal{R}_0$, and that this truncation may be approximated to a cut by a hyperplane for large $N$. The general integral $I_\mathcal{R}=\int_\mathcal{R}(\D\rvec{r}')\,f(\rvec{r}')$ is then simply a sum of values of the function $f$ in such an approximated truncated hyperellipsoid.

To simplify matters, we note that the truncated hyperellipsoid is mappable to a truncated hypersphere of some $\lambda$-dependent radius $R=R_\lambda$ in the diagonal basis of the covariance for the hyperellopsoid, so that we essentially have $I_\mathcal{R}\approx I_{S_{d-1}\backslash\mathrm{cap}_h}$, the truncated hyperspherical integral with a cap of height $0\leq h=h_\lambda\leq R$ removed. By invoking the hyperspherical coordinates and taking $f(\rvec{r}')=\rvec{r}'^2$ (the squared error with the center of the hypersphere assuming small statistical fluctuation $\rvec{r}\approx\ML{\rvec{r}}=\rvec{0}$), we may write
\begin{eqnarray}
I_{S_{d-1}\backslash\mathrm{cap}_h}&=&\frac{2\,\pi^{(d-1)/2} \,(1/2)!}{\left(d/2-1\right)!}\int_0^{\cos^{-1}(1-h/R)}\D\vartheta\,(\sin\vartheta)^{d-2}\nonumber\\
&&\qquad\qquad\qquad\qquad\times\left[\dfrac{R^{d+2}}{d+2}-\dfrac{1}{d+2}\left(\dfrac{R-h}{\cos\vartheta}\right)^{d+2}\right]
\end{eqnarray}
after some simple geometry. The truncated volume ($f(\rvec{r}')=1$) is also given by
\begin{eqnarray}
V_{S_{d-1}\backslash\mathrm{cap}_h}&=&\frac{2\,\pi^{(d-1)/2} \,(1/2)!}{\left(d/2-1\right)!}\int_0^{\cos^{-1}(1-h/R)}\D\vartheta\,(\sin\vartheta)^{d-2}\nonumber\\
&&\qquad\qquad\qquad\qquad\times\left[\dfrac{R^{d}}{d}-\dfrac{1}{d}\left(\dfrac{R-h}{\cos\vartheta}\right)^{d}\right]\,.
\end{eqnarray}
Altogether, we have $\mathrm{RSE}=\mathrm{RSE}_d\,(\rvec{r};h,R)=I_{S_{d-1}\backslash\mathrm{cap}_h}/V_{S_{d-1}\backslash\mathrm{cap}_h}$. 

The simple symmetry fact that $\mathrm{RSE}_d\,(\rvec{r};0,R)=\mathrm{RSE}_d\,(\rvec{r};R,R)$ forms the first key condition for the conservativeness proof of \eqref{eq:MRSE_exp2}. The second key condition for the proof is obtained by observing that the fraction of the integrand for $I_{S_{d-1}\backslash\mathrm{cap}_h}$ to that for $V_{S_{d-1}\backslash\mathrm{cap}_h}$ strictly increases with $\vartheta$. It therefore suffices to prove the following mathematical lemma for discrete summations and carry it over to integrations, which are also limited summations:\\[1ex]
\noindent
{\bf Lemma}---\emph{Let $\{a_j\geq0\}^N_{j=0}$ and $\{b_j\geq0\}^N_{j=0}$ for which $a_0/b_0=\sum^N_{j=0}a_j/\sum^N_{j=0}b_j$ and $a_j/b_j<a_{j+1}/b_{j+1}$ (strictly increasing fractions) for $1\leq j\leq N$. Then $\sum^k_{j=0}a_j/\sum^k_{j=0}b_j<a_0/b_0$ for $1\leq k\leq N-1$ and there is exactly one unique minimum value at $k=k^*$.}\\[1ex]
\noindent
{\bf Proof}---Define $t(k)\equiv\sum_{j=0}^{k}a_j/\sum_{j=0}^{k}b_j$ such that $a_0/b_0=t(0)=t(N)$. We first note that $a_0/b_0\geq(a_0+a_1)/(b_0+b_1)$. If not, $a_0/b_0<a_1/b_1<\cdots<a_N/b_N$ and we have eventually the inequality $t(k)<t(k+1)$ for $0\leq k\leq N-1$. This contradicts the initial condition $t(0)=t(N)$. Next, the fact that $\exists k=k^*\in[1,N-1]\,|\,t(k^*+1)\geq t(k^*)$ is obvious, and what remains is to show that $k^*$ is a unique point. This is straightforward since $t(k^*)\leq t(k^*+1)\implies t(k^*)\leq a_{k^*+1}/b_{k^*+1}$, and so the strictly increasing fraction chain then tells us that $t(k^*)\leq t(m)$ for $k^*<m\leq N$. \qed\\[1ex]

The above lemma implies that $\mathrm{RSE}\leq\mathrm{RSE}_\textsc{cat}=\mathrm{RSE}_d\,(\rvec{r};0,R)=\mathrm{RSE}_d\,(\rvec{r};R,R)$, which implies that the categorical RSE is an overestimate of the actual RSE for any $d$. The average sum of all these overestimates over all possible data then gives an overestimated MRSE. This is precisely the conservativeness property of \eqref{eq:MRSE_exp2}.

\section*{References}
\bibliography{Bayes.reg.2}

\end{document}